\documentclass[final,1p,dvipdfmx]{elsarticle}

\usepackage{amssymb}
\usepackage{color}

\journal{Physics Letters B}

\begin{document}

\begin{frontmatter}



\title{First measurement of coherent double neutral-pion photoproduction 
on the deuteron at incident energies below 0.9 GeV}


\author[elph]{T.~Ishikawa\corref{cor}}
\ead{ishikawa@lns.tohoku.ac.jp}
\author[elph]{H.~Fujimura\fnref{hf}}
\author[elph]{H.~Fukasawa}
\author[elph]{R.~Hashimoto\fnref{rh}}
\author[elph]{Q.~He}
\author[elph]{Y.~Honda}
\author[yama]{T.~Iwata}
\author[elph]{S.~Kaida}
\author[aoba]{H.~Kanda}
\author[elph]{J.~Kasagi}
\author[gaku]{A.~Kawano}
\author[elph]{S.~Kuwasaki}
\author[aoba]{K.~Maeda}
\author[tokyo]{S.~Masumoto}
\author[elph]{M.~Miyabe}
\author[elph]{F.~Miyahara\fnref{fm}}
\author[elph]{K.~Mochizuki}
\author[elph]{N.~Muramatsu}
\author[elph]{A.~Nakamura}
\author[elph]{K.~Nawa}
\author[elph]{S.~Ogushi}
\author[elph]{Y.~Okada}
\author[elph]{K.~Okamura}
\author[elph]{Y.~Onodera}
\author[kek]{K.~Ozawa}
\author[gaku]{Y.~Sakamoto}
\author[elph]{M.~Sato}
\author[elph]{H.~Shimizu}
\author[elph]{H.~Sugai\fnref{hs}}
\author[elph]{K.~Suzuki\fnref{ks}}
\author[yama]{Y.~Tajima}
\author[elph]{Y.~Taniguchi}
\author[elph]{Y.~Tsuchikawa\fnref{yt}}
\author[elph]{H.~Yamazaki\fnref{hy}}
\author[elph]{R.~Yamazaki}
\author[yama]{H.Y.~Yoshida}
\address[elph]{Research Center for Electron Photon Science (ELPH), Tohoku University, Sendai 982-0826, Japan}
\address[yama]{Department of Physics, Yamagata University, Yamagata 990-8560, Japan}
\address[aoba]{Department of Physics, Tohoku University, Sendai 980-8578, Japan}
\address[gaku]{Department of Information Science, Tohoku Gakuin University, Sendai 981-3193, Japan}
\address[tokyo]{Department of Physics, University of Tokyo, Tokyo 113-0033, Japan}
\address[kek]{Institute of Particle and Nuclear Studies, High Energy Accelerator Research Organization (KEK), Tsukuba 305-0801, Japan}

\cortext[cor]{Corresponding author. Tel.: +81 22 743 3400; fax: +81 22 743 3401.}
\fntext[hf]{Present address: Department of Physics, Wakayama Medical University, Wakayama 641-8509, Japan}
\fntext[rh]{Present address: Institute of Materials Structure Science (IMSS), High Energy Accelerator Research Organization (KEK), Tsukuba 305-0801, Japan}
\fntext[fm]{Present address: Accelerator Laboratory, High Energy Accelerator Research Organization (KEK), Tsukuba 305-0801, Japan}
\fntext[hs]{Present address: Gunma University Initiative for Advanced Research (GIAR), Maebashi 371-8511, Japan}
\fntext[ks]{Present address: The Wakasa Wan Energy Research Center, Tsuruga 914-0192, Japan}
\fntext[yt]{Present address: Department of Physics, Nagoya University, Nagoya 464-8602, Japan}
\fntext[hy]{Present address: Radiation Science Center, High Energy Accelerator Research Organization (KEK), Tokai 319-1195, Japan}
\begin{abstract}
The total cross sections 
were measured for coherent double neutral-pion photoproduction on the deuteron
at incident energies below 0.9 GeV for the first time.
No clear resonance-like behavior is observed in the excitation function
for $W_{\gamma d}=2.38$--2.61 GeV,
where the $d^*(2380)$ dibaryon resonance observed at COSY is expected 
to appear.
The measured excitation function is consistent with the existing theoretical 
calculation for this reaction.
The upper limit of the total cross section is found to be 
$0.034$~$\mu$b
for the dibaryon resonance at $W_{\gamma d}=2.37$~GeV (90\% confidence level) in the $\gamma d \to \pi^0\pi^0 d$ reaction.
\end{abstract}


\end{frontmatter}

The internal structure of hadrons is a subject in the non-perturbative domain of the 
fundamental theory of strong interactions, quantum chromodynamics.
The familiar mesons and baryons are composed of $q\bar{q}$ and $qqq$, respectively.
More complex quark configurations beyond these are objects of great interest 
to investigate the effective degrees of freedom describing hadrons
and to understand color confinement.
The WASA-at-COSY collaboration has recently reported the isoscalar $d^*(2380)$ resonance with 
mass $M\simeq 2380$~MeV and width $\Gamma\simeq 68$~MeV,
which is observed in the $pn\to \pi^0\pi^0d$~\cite{cosy1} and $pn\to \pi^+\pi^- d$~\cite{cosy2} reactions.
The first indication corresponding to this resonance was observed 
in the former reaction by the CELSIUS/WASA collaboration~\cite{cels}.
The resonance may be attributed to an isoscalar $\Delta\Delta$ quasi-bound state, ${\cal D}_{03}$,
predicted by Dyson and Xuong~\cite{dx}.
In addition to the $\pi^0 \pi^0d$ and  $\pi^+\pi^-d$ final states,
evidence for the $d^*(2380)$ resonance has been confirmed
by the WASA-at-COSY collaboration in the $\pi^0\pi^-pp$~\cite{cosy3},
$\pi^0\pi^0pn$~\cite{cosy4}, and $\pi^+\pi^-pn$~\cite{cosy5} final states.
The SAID partial wave analysis,
which incorporates the analyzing power for the quasi-elastic $\vec{n}p\to np$ scattering
measured by the WASA-at-COSY collaboration,
also supports the existence of the $d^*(2380)$
resonance with quantum numbers $I(J^\pi)=0(3^+)$~\cite{cosy6,cosy7}.
These experimental results have 
stimulated intensive theoretical investigations 
of  ${\cal D}_{03}$~\cite{ds1,ds2}.
To date, all the observations have been made using $pn$ collisions.
Nearly all the measurements were made by the WASA-at-COSY collaboration.

The $d^*(2380)$ resonance should be observable in photoproduction reactions
if it exists.
The $\gamma d\to \pi^+\pi^-d$ and $\gamma d\to \pi^0\pi^0d$ reactions 
are expected to be of value when studying the production mechanism of 
the $d^*(2380)$ resonance.
It may be produced as an intermediate state in the $s$ channel,
and decays into a final state including a deuteron, where
no special treatment is required kinematically  for 
the Fermi motion of nucleons.
The $\pi^+\pi^-d$ final state includes the isovector ($I  = 1$) component,
while the $\pi^0\pi^0d$ has just the isoscalar ($I = 0$) component alone.
The Kroll-Ruderman contact term is expected to give a large effect 
in the $\pi^+\pi^-d$ channel, i.e., the $\gamma N\pi^\pm$ coupling is large.
This may hide the $d^*(2380)$ contribution in the $\gamma d \to \pi^+\pi^-d $ yield.
This reaction was studied by the CLAS collaboration
at the Thomas Jefferson National Accelerator Facility. Their preliminary result
does not show a peak corresponding to the $d^*(2380)$ resonance~\cite{jlab}.
The other
$\gamma d\to \pi^0\pi^0d$ reaction is thought to be the best process 
to investigate the $d^*(2380)$ resonance in photoproduction.

Two series of meson photoproduction experiments~\cite{exp}
were carried out using the tagged photon beam~\cite{tag2,bpm}
at the Research Center for Electron Photon Science (ELPH), Tohoku University.
The photon beam was produced by a bremsstrahlung process with a carbon fiber
from the 0.93 GeV circulating electrons in a synchrotron called the STretcher Booster (STB) ring~\cite{stb}.
The tagging energy of the photon beam ranged from 0.57 to 0.88 GeV.
The target used in the experiments was 
liquid deuterium with a thickness of 45.9~mm. 
The incident photon energy gives a $\gamma d$ center of mass energy,
$W_{\gamma d}$, from 2.38 to 2.61 GeV,
and the lowest photon energy corresponds to the centroid of the $d^*(2380)$ resonance.

All the final-state particles in the $\gamma d \to \pi^0\pi^0d $ reaction 
were measured using an electromagnetic (EM) 
calorimeter complex, FOREST~\cite{forest}.
FOREST consists of three different EM calorimeters:
the forward, central, and backward calorimeters consisting of 
192 pure CsI crystals, 252 lead scintillating fiber modules, and
62 lead glass Cherenkov counters, respectively.
A plastic scintillator (PS) hodoscope is placed in front of each calorimeter
to identify the charged particles.
The solid angle of FOREST is approximately 88\% in total.
The typical tagging rate was 2.8~MHz,
and the photon transmittance (so-called tagging efficiency)
was approximately 42\%~\cite{tag2}.
The trigger condition of the data acquisition (DAQ)
was
\begin{equation}
\sum_i [{\rm ST\ } i]\otimes [\#{\rm S3}+\#{\rm BG}\ge 2]
\end{equation}
to detect more than one final-state particles in coincidence with a tagging signal,
where $\sum$ indicates the OR signal of signals and 
$\otimes$ stands for  the coincidence of signals.
The $\#{\rm S3}+\#{\rm BG} \ge 2$ denotes the signal generated 
when two output signals out of the groups in the forward and central
calorimeters were given.
The ST $i$ denotes an OR signal of the
tagging channels in the corresponding group $i=1,\ldots,16$.
The details of the groups are described elsewhere~\cite{forest}.
The average trigger rate was 1.1~kHz, 
and the average DAQ efficiency was 85\%.

Events detected in the final state containing 
four neutral particles and a charged particle 
were selected.
Each neutral pion in the $\gamma d\to \pi^0\pi^0d$ reaction was identified 
via its decay into $\gamma\gamma$. 
Photons were detected as a set (cluster) of hit calorimeter modules
without any responses of the hit PSs in the front hodoscope.
The details of making clusters in FOREST are described in Ref.~\cite{forest}.
The time difference between every two neutral clusters of four 
was required to be less than $3\sigma_t$, where $\sigma_t$ denotes 
the time resolution for the difference depending on the modules and their
measured energies for the two clusters.
Deuterons in the final state were detected with the forward hodoscope called SPIDER, and the direction of emission was determined 
by the hit PSs. Note that the response of the corresponding 
calorimeter called SCISSORS III was not required.
The time delay from the average time response between the four neutral 
clusters was required to be larger than 1 ns.
The energy measured with SPIDER was required to be greater
than $2E_{\rm mip}$, where $E_{\rm mip}$ denotes the energy that the minimum 
ionizing particle deposits in a PS.
The momentum of deuterons was calculated from the measured time delay
assuming that the charged particles had the mass of the deuteron.

A kinematic fit with six constraints (6C) was applied for the further 
event selection of the 
$\gamma d\to \pi^0\pi^0d$ reaction.
The kinematic variables in the fit were the incident photon energy,
the three-momentum of the five final-state particles, and
the reaction vertex point.
Even though FOREST did not have a vertex counter, 
the $(x,y)$ intensity map of the photon beam was measured using
a beam-profile monitor~\cite{bpm} day by day.
The measured variable and its resolution for the $x$($y$)-component 
of the vertex point were assumed to be the same as the centroid and width
of the $x$($y$) distribution of the photon beam at the target position.
Because the attenuation of the photon beam flux was negligibly small
passing through the liquid deuterium target,
the measured variable and its resolution for the $z$-component was
assumed  to be the same as the center and thickness($\sigma$) of the target.
The required constraints were 
energy and three-momentum conservation between the initial and final states
and two $\gamma\gamma$ invariant masses (the neutral-pion rest mass, $m_{\pi^0}$).

The 6C kinematic fit is effective at selecting the 
$\gamma d \to \pi^0\pi^0 d$ reaction.
Events in which the $\chi^2$ probability was higher
than 0.4 were selected to prevent contamination 
from the quasi-free two neutral-pion photoproduction 
on the proton in the deuteron, $\gamma p'\to \pi^0\pi^0p$.
This quasi-free production is the most competitive
background process, having 100 times higher cross 
section~\cite{kurs}.
The lower limit of $\chi^2$ probability 0.4 makes 
the contamination less than 5\%, which is much less 
than the statistical error of the measured total 
cross section ($\sim 20\%$).
Because accidental coincidence events exist
between the photon-tagging counter, STB-Tagger II~\cite{tag2}, and FOREST, 
sideband background subtraction was performed.

\begin{figure}[htb]
\begin{center}
\includegraphics[width=0.9\textwidth]{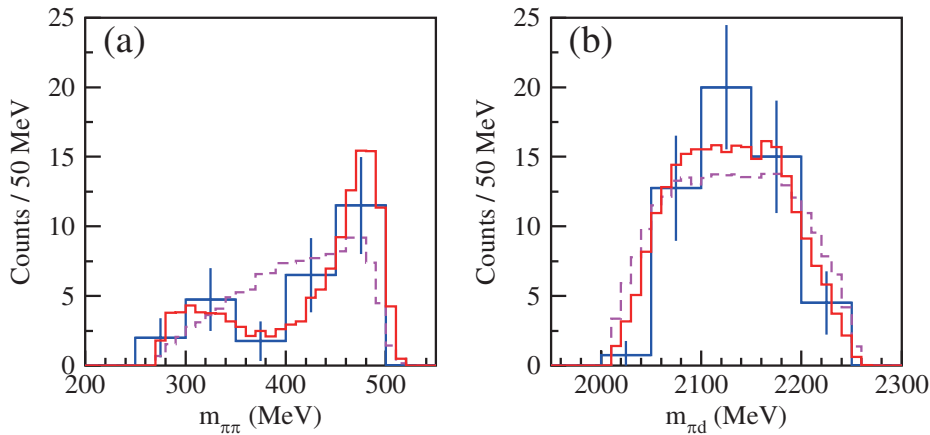}
\end{center}
\caption{(a) $\pi^0\pi^0$ 
and (b) $\pi^0d$ invariant mass distributions at $W=2.39$~GeV 
(2.382--2.396 GeV).
The data points (blue) are compared with the simulation results.
The dashed histogram (magenta) shows the results 
for pure phase-space event generation,
and the solid histogram (red) shows the results for $n=4.9$ (see text).
Normalizations of the simulation results are the same for  (a) and (b).
}
\label{fig2}
\end{figure}

An invariant mass distribution of two final-state
particles was investigated to give the difference
of the measured distributions between the experimental
data and pure phase-space simulation.
Fig.~\ref{fig2}(a) shows the typical $\pi^0\pi^0$ 
invariant mass ($m_{\pi\pi}$)  distribution.
The $m_{\pi\pi}$ distribution for the real data 
is quite different from that for the pure phase-space
generation of the three final-state particles.
An enhancement is observed in the lower-mass region
close to $2m_{\pi^0}$, which may correspond to 
the ABC effect~\cite{abc}.
In addition, another enhancement is observed 
in the higher-mass region.
These two enhancements are observed in all the 
incident energy regions.
Fig.~\ref{fig2}(b) shows the typical $\pi^0d$ invariant 
mass  ($m_{\pi d}$)  distribution.
No significant difference between the real data and 
the simulation is observed in the $m_{\pi d}$ distribution.

The total cross section of the $\gamma d \to \pi^0\pi^0 d$ reaction 
can be obtained from the equation
\begin{equation}
\sigma = \frac{
N_{\pi^0\pi^0d}
}{
N'_\gamma
N_\tau
\eta_{\rm acc}
\left\{
{\rm BR}({\pi^0\to \gamma\gamma})\right\}^2\label{eq:1}
},\end{equation}
which uses 
the number of events for the $\gamma d\to\pi^0\pi^0 d$ reaction, $N_{\pi^0\pi^0d}$,
the effective number of incident photons, $N'_\gamma$,
the number of target deuterons per unit area, $N_\tau=0.237$ b${}^{-1}$,
the acceptance of the final state $\pi^0\pi^0 d\to \gamma\gamma\gamma\gamma d$ detection,
$\eta_{\rm acc}$, and the branching ratio of the neutral pion 
to the two-photon decay, ${\rm BR}({\pi^0\to \gamma\gamma})$.
The number of incident photons, $N_\gamma$, is determined by
multiplying the number of tagging signals after the counting-loss correction
by the corresponding photon transmittance.
The $N'_\gamma$ is obtained additionally multiplying $N_\gamma$ by the DAQ efficiency.
The acceptance of  $\gamma\gamma\gamma\gamma d$ detection
is estimated by a Monte-Carlo simulation based on Geant4~\cite{geant4}.
Here, the total cross section as a function of the incident energy is assumed to be flat.
To reproduce the measured $m_{\pi\pi}$ distribution,
the $m_{\pi\pi}$ distribution for generated events is assumed to have
an additional dependence from pure phase-space generation:
\begin{equation}
P =	
\left(\frac{m_{\pi\pi}-m_{\pi\pi}^{\rm min}}
{m_{\pi\pi}^{\rm max}-m_{\pi\pi}^{\rm min}}\right)^n
+
\left(\frac{m_{\pi\pi}^{\rm max}-m_{\pi\pi}}
{m_{\pi\pi}^{\rm max}-m_{\pi\pi}^{\rm min}}\right)^n
\label{eq2}
\end{equation}
with $n=4.9$, where 
$m_{\pi\pi}^{\rm max}$ and 
$m_{\pi\pi}^{\rm min}$ denote the maximum and minimum values for $m_{\pi\pi}$,
respectively, at the fixed incident photon energy.

\begin{figure}[htb]
\begin{center}
\includegraphics[width=0.9\textwidth]{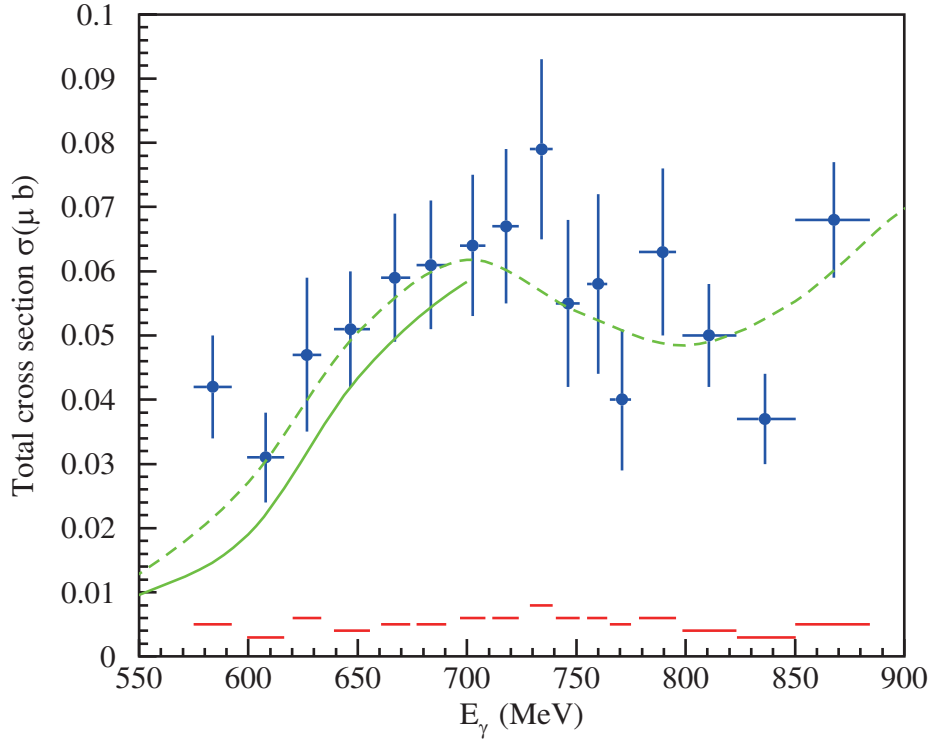}
\end{center}
\caption{Total cross section, $\sigma$, as a function of  $E_{\gamma}$. 
The upper points (blue) show the obtained $\sigma$.
The horizontal bar of each point shows the coverage of the incident photon 
energy, and the vertical bar shows the statistical error of $\sigma$.
The lower histogram (red) shows the systematic errors (see text for details).
The data are compared with theoretical calculations
for the $\gamma d\to \pi^0\pi^0 d$ reaction 
given in Ref.~\cite{fix1} (dashed) and Ref.~\cite{fix2} (solid).
}
\label{fig4}
\end{figure}

Because the statistics were limited, the tagging channels were divided into 16 groups,
and the total cross section was obtained for each group.
Fig.~\ref{fig4} shows the total cross section, $\sigma$, for the 
$\gamma d\to\pi^0\pi^0 d$ reaction,
as a function of  the incident energy, $E_{\gamma}$. 
The total cross section is rather flat, and a 
clear resonance-like behavior is not 
observed in the excitation function
for $E_\gamma = 0.57$--$0.88$~GeV ($W_{\gamma d}=2.38$--2.61 GeV).

To estimate the systematic uncertainty of event yields,
we varied the lower limit of event selection in the kinematic fit from 0.2 to 0.6
(from 19\% to 3\% contamination assuming 100 times higher total cross section
for the quasi-free $\gamma p' \to \pi^0\pi^0p$ reaction),
and the uncertainty ($\sigma$) was found to range from 5.2\% to 10.8\%
depending on the tagging-energy group.
To estimate the systematic uncertainty of the acceptance,
we changed the $m_{\pi\pi}$ distribution for event generation in the simulation.
The $n$  parameters corresponding to the realistic $m_{\pi\pi}$ distributions
give the uncertainty ($\sigma$) from 0.1\% to 0.5\%.
The uncertainty in the acceptance from the uncertainty in the FOREST coverage is 
0.7\%--4.1\%.
The uncertainty in the deuteron detection efficiency is 1.0\%--5.3\%
owing to the uncertainty in the density of the vacuum chamber 
surrounding the liquid deuterium target.
The normalization uncertainties resulting from the 
number of target deuterons and the number of incident photons 
are 1\% and 1.5\%--1.9\%, respectively.
The total systematic uncertainty is obtained by combining all the uncertainties 
described above in quadrature. 
The total systematic uncertainty as a function of $E_\gamma$ is 
also plotted in Fig.~\ref{fig4}.

Fix and Arenh\"ovel reported their calculation of the total cross section
for the $\gamma d \to \pi^0\pi^0 d$ reaction at $E_\gamma=0.32$--1.50 
GeV~\cite{fix1}.
Egorov and Fix recently reported their calculation for the reaction
at $E_\gamma=0.40$--0.70 GeV~\cite{fix2}. 
The calculated cross sections are also plotted in Fig.~\ref{fig4}.
For coherent production,
the isovector parts in the amplitudes for the $\pi^0\pi^0$ production
on the proton and neutron are canceled.
Because the fraction of the isoscalar part is thought to be
only 8\% of the proton amplitude,
the cross section for the coherent production is much smaller than
that for the quasi-free $\pi^0\pi^0$ production on the 
nucleon ($\sim 10$~$\mu$b at $E_\gamma$=0.60 GeV).
The measured cross section is well reproduced by 
the calculation given by Fix and Arenh\"ovel except 
for the lowest incident photon energy region ($\sim$0.57~GeV).
The discrepancy in the lowest energy region 
may be explained by excitation of the $d^*(2380)$ dibaryon 
resonance.

\begin{figure}[htb]
\begin{center}
\includegraphics[width=0.9\textwidth]{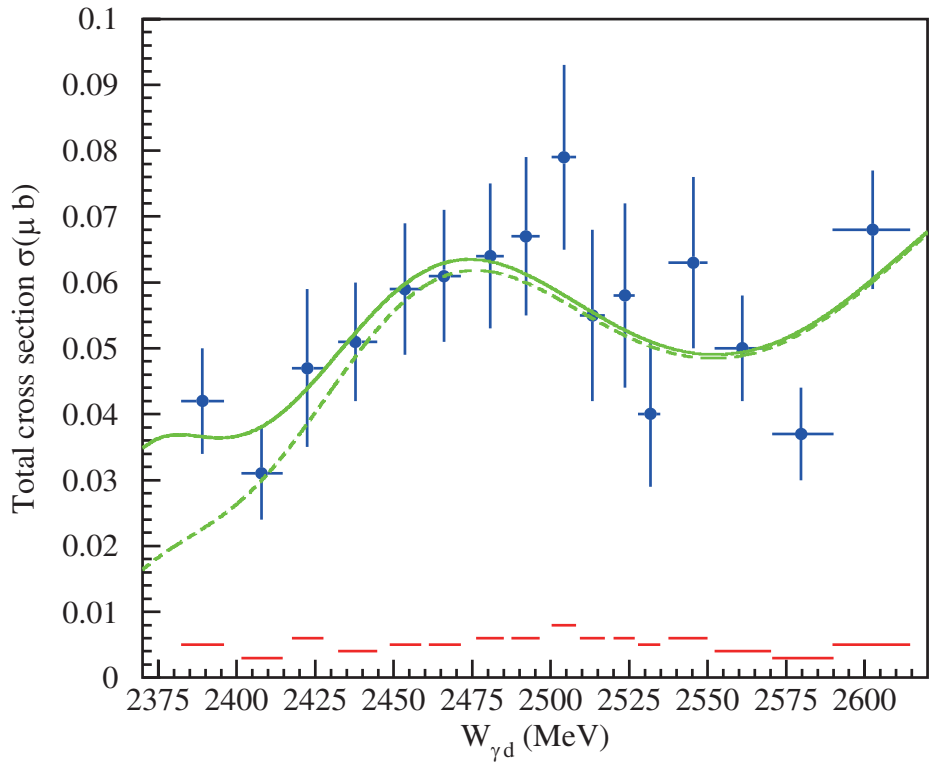}
\end{center}
\caption{
Total cross section, $\sigma$, as a function of  $W_{\gamma d}$. 
The upper points (blue) show the obtained $\sigma$.
The horizontal error of each point corresponds to 
the coverage of the incident photon energy, and the vertical error shows 
the statistical error of $\sigma$.
The lower histogram (red) shows the systematic error of $\sigma$ (see text for details).
The dotted curve (green) shows the calculated $\sigma$ given in 
Ref.~\cite{fix1}.
The data are compared with a function shown in the solid curve (green)
expressed by the sum of the expected $d^*(2380)$ contribution
with a relativistic Breit-Wigner 
shape with $W=2.37$~GeV and $\Gamma=68$~MeV
(0.0184 $\mu$b at $W=2.37$ GeV) and calculated $\sigma$~\cite{fix1}.
}
\label{fig5}
\end{figure}

Fig.~\ref{fig5} shows the total cross section, $\sigma$, for the 
$\gamma d\to\pi^0\pi^0 d$ reaction as a function of $W_{\gamma d}$.
The $d^*(2380)$ contribution was estimated by fitting the function
\begin{equation}
\sigma(W_{\gamma d}) =
\frac{{\rm BW}(W_{\gamma d})}{
{\rm BW}(2.37{\rm\ GeV})}\sigma_{d^*}
+ \sigma_{\rm th}(W_{\gamma d})
\end{equation}
to the data,
where BW$(W_{\gamma d})$ denotes the relativistic Breit-Wigner function~\cite{bw}
corresponding to the expected  $d^*(2380)$ contribution 
with a centroid of $M=2.37$~GeV and a width of $\Gamma=68$~MeV.
The $\sigma_{\rm th}$ stands for the calculated cross section 
given by Fix and Arenh\"ovel.
The $\chi^2/{\rm dof}$ of the fit is 10.1/15, and the obtained parameter is 
\begin{equation}
\sigma_{d^*} = 0.0184\pm 0.0091 {\rm\ }\mu{\rm b}.
\end{equation}
The upper limit of the total cross section was found to be $0.034$~$\mu$b
at $W_{\gamma d}=2.37$~GeV (90\% confidence level).

The total cross sections for the $\gamma d \to \pi^0\pi^0 d$ reaction
were measured for the first time using the FOREST detector at ELPH.
The incident energy ranged from 0.57 to 0.88  GeV.
No clear resonance-like behavior
corresponding to the $d^*(2380)$ resonance with $I(J^\pi)=0(3^+)$
was observed in the excitation function for $W_{\gamma d}=2.38$--2.61~GeV.
The measured cross section is well reproduced by 
the calculation given by Fix and Arenh\"ovel~\cite{fix1} except 
for the lowest incident photon energy region ($\sim$0.57~GeV).
A possible explanation of the discrepancy in the lowest energy region
may be attributed to excitation of the $d^*(2380)$ dibaryon 
resonance.
The upper limit of the total cross section in this reaction was found to 
be $0.034$~$\mu$b for the dibaryon resonance at $W_{\gamma d}=2.37$~GeV (90\% confidence level).
A further understanding of the $\gamma d\to \pi^0\pi^0d$ 
reaction mechanism is required.

The authors express their gratitude to the ELPH accelerator staff for stable operation 
of the accelerators in the FOREST experiments.
In addition, they acknowledge Mr.\ K.~Matsuda for his excellent technical assistance 
for the construction of FOREST including the preparation of the 
cryogenic target system.
The authors are grateful to Mr.~K.~Nanbu, and Mr.~I.~Nagasawa for 
their technical support in the preparation of the data acquisition system
for the FOREST experiments.
The authors also thank the Yukawa Institute for Theoretical Physics (YITP), 
Kyoto University.
Discussions during the workshop YITP-W-16-01 ``Meson in Nucleus 2016 (MIN16)''
were useful in completing this work.
We also thank Prof.\ A.~Fix for providing numerical values of the total cross section. This work was supported in part by the Ministry of Education, Culture, Sports, Science and Technology, Japan
through Grants-in-Aid for Scientific Research (B) No.\ 17340063, 
for Specially Promoted Research No.\ 19002003,
for Scientific Research (A) No.\ 24244022,
for Scientific Research (C) No.\ 26400287,
and for Scientific Research (A) No.\ 16H02188.


\end{document}